\documentclass[a4paper,
preprint,
preprintnumbers,
nofootinbib,
 amsmath,amssymb,
]{revtex4-1}

\usepackage{cancel}
\usepackage{graphicx}
\usepackage{tabularx}
\usepackage{colortbl}
\usepackage{dcolumn}
\usepackage{bm}
\usepackage{latexsym}
\usepackage{amsmath,amsfonts,amssymb}
\usepackage{graphicx,epsfig}
\usepackage{subfig}
\usepackage{psfrag}
\usepackage{pgfplots}
\usepackage{epsfig}
\usepackage{epstopdf}
\usepackage{psfrag}
\usepackage{hyperref}
\usepackage[mathlines]{lineno}
\usepackage{tikz} 
\usepackage{tikz-feynman}
\tikzfeynmanset{compat=1.1.0}
\usepackage{color}
\usepackage{braket}
\usepackage[english]{babel}
\usepackage{array}

\definecolor{dgreen}{rgb}{0,0.39,0}

\begin{document}
\title{Quantum field theory with the generalized uncertainty principle I: scalar electrodynamics}

\author{Pasquale Bosso}
\email{pasquale.bosso@uleth.ca}
\affiliation{Theoretical Physics Group and Quantum Alberta,
Department of Physics and Astronomy,
University of Lethbridge,
4401 University Drive, Lethbridge,
Alberta, T1K 3M4, Canada}

\author{Saurya Das}
 \email{saurya.das@uleth.ca}
\affiliation{Theoretical Physics Group and Quantum Alberta,
Department of Physics and Astronomy,
University of Lethbridge,
4401 University Drive, Lethbridge,
Alberta, T1K 3M4, Canada}

\author{Vasil Todorinov}
 \email{v.todorinov@uleth.ca}
\affiliation{Theoretical Physics Group and Quantum Alberta,
Department of Physics and Astronomy,
University of Lethbridge,
4401 University Drive, Lethbridge,
Alberta, T1K 3M4, Canada}

 
\date{\today}

%
%
%
%

\begin{abstract}
Recently, the authors presented a covariant extension of the Generalized Uncertainty Principle (GUP) with a 
Lorentz invariant minimum length.
This opens the way for constructing and exploring the observable 
consequences of minimum length in Relativistic Quantum Field Theories.
In particular, we compute 
quantum gravity corrections to high energy scattering experiments, which may provide the much needed window of testing minimum length and quantum gravity theories in the laboratory. 
To this end, we formulate the Lagrangian of Quantum Electrodynamics  for a complex scalar fields from the GUP modified minimally coupled 
Klein-Gordon equation and write down its Feynman rules.
We then calculate the Relativistic Generalized Uncertainty Principle corrections to a Quantum Electrodynamics scattering amplitude and discuss its implications. 
\end{abstract}
\pacs{Valid PACS appear here}
%
\maketitle

\newpage

\section{\label{sec:Int}Introduction}

A common prediction of theories of Quantum Gravity is the existence of a minimum measurable length
in spacetime \cite{Physics1988,AHARONY2000183,Magueijo:2004vv,Rovelli:2010vv,Rovelli:2010wq,Smolin:2004sx,Antonio2018,Amelino-Camelia2010,Smolin2003}.
Phenomenological implications of this prediction are generally studied 
using the Generalized Uncertainty Principle (GUP) \cite{Kempf1995,Ali2011,Bosso2018,Ali2010,Das2011,Das2008,Das2014}.
However, apart from a few notable exceptions \cite{Hossenfelder2006,Pramanik2013_1,Deriglazov2014,Pramanik2014_1,Capozziello:1999wx,Kober:2010sj,Kober:2011dn,Shibusa:2007ju,Husain:2012im,Zakrzewski1994,Hill2002,Faizal:2014dua,Faizal:2017map,Szabo:2006wx,Chaichian:2004za,Snyder:1946qz,Quesne2006}, the majority of those studies have been in the context on non-relativistic quantum mechanics.
In a previous work, we extended GUP to the relativistic framework \cite{Todorinov:2018arx}.
In addition, we showed that this  Relativistic Generalized Uncertainty Principle (RGUP) yields a Lorentz invariant minimum length.
This extension led to some interesting properties of spacetime, such as its non-commutativity at high energies.
Furthermore, a relationship between the parameters of the RGUP was found, allowing the non-commutative spacetime to preserve the symmetries of the classical spacetime.
Thus, with a fully consistent and relativistic GUP and minimum length, it is natural to explore the implications for Relativistic Quantum Field Theory, given especially that quantum gravity effects are expected to become relevant at high energies.
This is the primary motivation of the present work. 

We consider a quadratic form for RGUP, that is
\begin{equation}
\label{GUPxp}[x^{\mu},p^{\nu}]=i\hbar\,\left(1-\gamma  p^{\rho}p_{\rho}\right)\eta^{\mu\nu}+i\,\hbar\gamma p^{\mu}p^{\nu}\,,
\end{equation}
where $\gamma=\frac{\gamma_0}{(M_{\mathrm{Pl}}\,c)^2}$, $\gamma_0$ is a dimensionless parameter used to set the scale for the quantum gravitational effects.
For the purposes of this work, we will consider all physical quantities in natural units, \emph{i.e.} $\hbar=c=1$.
In this units, we have $\gamma =\gamma_0 \ell_{\mathrm{Pl}}^2$.
From the relation above, we can easily see that position and momentum are no longer canonically conjugate.
Therefore, we introduce a pair of new canonically conjugated auxiliary  variables, $x_0^\mu$ and $p_0^\mu$, such that
\begin{align}
\label{MathVar}p_0^{\mu} = & -i\frac{\partial}{\partial x_{0\,\mu}}, &
[x_0^{\mu},p_0^{\nu}] = & i\eta^{\mu\nu}\,.
\end{align}
Using these new variables, we express the position and momentum operators in terms of canonically conjugate variables.
The definition of these auxiliary quantities are useful in deriving the modification of the spacetime symmetries, \emph{i.e.} the Poincar\'e group, resulting from Eq.\eqref{GUPxp} in \cite{Todorinov:2018arx}.
For the purpose of the present work, we will consider the auxiliary and physical positions to be equivalent, $x^\mu=x^\mu_0$.
We can prove that the squared physical momentum $p^{\rho}p_{\rho}$ is again a Casimir invariant of the modified Poincar\'e group.
In other words, it commutes with every other operator in the group.
Furthermore, using it we can derive the Klein-Gordon (KG) equation.
In terms of the auxiliary variables $p_0^{\mu}$, we have
\begin{equation}
    \label{ModKG}
    p_0^{\rho}p_{0\rho}(1+2\gamma p_0^{\sigma}p_{0\sigma}
    )=-(mc)^2 \,.
\end{equation}

In this work we explore the consequences of minimum length represented by RGUP modified dispersion relation Eq.\eqref{ModKG} and its effects on Relativistic Quantum Field Theories. We organized our results as follows:
in Section \ref{sec:Lag}, we reconstruct the Lagrangians for the RGUP modified scalar field theory using the Ostrogradsky method to work with higher derivative equations of motion.
Then, we proceed to derive the Feynman propagators for both gauge and scalar fields, followed by the introduction of minimal coupling and the derivation of the Feynman rules.
In Section \ref{sec:Crosssection}, we derive RGUP corrections for the amplitude for scattering of a scalar muon and electron. Furthermore, we calculate the cross-section of the scattering in the 
ultra-relativistic regime.
We summarise our results and discus future directions in Section \ref{sec:Conclusion}. 
\\

\section{\label{sec:Lag}Scalar field Lagrangian}

Let us first consider a free massive scalar field with the following equation of motion
\begin{equation}
\left[p^{\mu}p_{\mu} + m^2\right]\phi=0\,.
\end{equation}
For simplicity, let us consider $1+1$ dimensions.
The equation above then has the following form 
\begin{equation}\label{KG1+1}
    \left[p_{t}^2-p_{x}^2 + m^2\right]\phi=0\,.
\end{equation}
Expressing the physical momentum in terms of the auxiliary variables
\begin{align}
    p_{t} = & p_{0\,t}\left[1+\gamma \left(p_{0\,t}^2-p_{0\,x}^2\right)\right], &
    p_{x} = & p_{0\,x}\left[1+\gamma \left(p_{0\,t}^2-p_{0\,x}^2\right)\right]\,,
\end{align}

Eq.\eqref{KG1+1} can be written as 
\begin{equation}
\left\{\left(p_{0\,t}^2-p_{0\,x}^2\right)\left[1+2\gamma\left(p_{0\,t}^2-p_{0\,x}^
2\right)\right]+ m^2\right\}\phi = 0 \,.
\label{EoMp0}
\end{equation}
Introducing the operator $\Box =\frac{\partial}{\partial x_{\mu}}\frac{\partial}{\partial x^\mu}$, we generalize  Eq.\eqref{EoMp0} to $(3+1)$-dimensions as follows 
\begin{equation}\label{EoM}
    \left(\Box +2\gamma \Box^2
    + m^2 \right)\phi=0\,.
\end{equation}
\subsection{Quantum gravity modified Lagrangians}
Using the Ostrogradsky method for higher derivative Lagrangians \cite{Pons:1988tj,Woodard:2015zca,deUrries:1998obu}, we obtains the following Lagrangian (see Appendix \ref{apx})
\begin{equation}\label{RealL}
    \mathcal{L}_{\phi,\mathbb{R}}=\frac{1}{2}\partial_{\mu}\phi\partial^{\mu}\phi-\frac{1}{2}m^2\phi^2+\gamma \,\partial_\nu\partial^\nu\partial^\mu\phi\,\partial_\mu\phi
    \,,
\end{equation}
where $\partial_\mu = \partial / \partial x^\mu$.
As for the Lagrangian for a complex scalar field $\phi$, we generalize Eq.\eqref{RealL} by including additional terms obtaining 
\begin{equation}\label{complexL}
 \mathcal{L}_{\phi,\mathbb{C}} = \frac{1}{2} \left(\partial_{\mu}\phi\right)^\dagger \partial^{\mu}\phi - \frac{1}{2} m^2 \phi^\dagger \phi + \gamma \left[\left(\partial_\nu \partial^\nu \partial^\mu \phi \right)^\dagger \partial_\mu \phi + \partial_\nu \partial^\nu \partial^\mu \phi \left(\partial_\mu \phi \right)^\dagger \right] \,,
\end{equation}
such that hermiticity is restored, \textit{i.e.} $\mathcal{L}_{\phi,\mathbb{C}}^{\dagger}=\mathcal{L}_{\phi,\mathbb{C}}$.
Furthermore, it is worth noticing that Eq.\eqref{complexL} is consistent with Eq.(55) in \cite{Kober:2011dn} up to a numerical factor.

We assume that Electrodynamics Lagrangian to have the standard spacetime and $U(1)$ gauge symmetries. As for the equations of motion, we assume they have the same GUP corrections as the KG equation, of the form 
 \begin{equation}\label{gaugefieldEOM}
     \partial_\mu F^{\mu\nu}= \partial_\mu\partial^\mu A^\sigma +2\gamma \partial_\mu\partial^\mu\partial_\nu\partial^\nu A^\sigma  
     =0\,.
 \end{equation}
 Defining the standard gauge invariant
 field strength tensor 
 \begin{equation}
     F^{\mu\nu}_0=\partial^\mu A^\nu - \partial^\nu A^\mu,
 \end{equation}
 We can express the RGUP modified field strength tensor in terms of the standard one up to first order in $\gamma$, as follows:
 \begin{equation}
    F^{\mu\nu}=F^{\mu\nu}_0+2\gamma\, \partial_\rho \partial^\rho F^{\mu\nu}_0\,.
 \end{equation}
Then Eq.\eqref{gaugefieldEOM} can be rewritten as 
 \begin{equation}\label{fieldtensor}
    \partial_\mu F^{\mu\nu}=\partial_\mu F^{\mu\nu}_0+2\gamma  \partial_\rho \partial^\rho\partial_\mu F^{\mu\nu}_0
    \,.
 \end{equation}
 Thus, the gauge field Lagrangian reads
 \begin{equation}\label{GaugeFieldL}
     \mathcal{L}_{A}=-\frac{1}{4}F^{\mu\nu}F_{\mu\nu}=-\frac{1}{4}F^{\mu\nu}_0F_{\mu\nu 0}-\frac{\gamma }{2}F_{\mu\nu 0} \partial_\rho \partial^\rho F^{\mu\nu}_0
     \,.
 \end{equation}
Notice that both the field tensor in Eq.\eqref{fieldtensor} and the gauge field Lagrangian Eq.\eqref{GaugeFieldL} are gauge invariant.

\subsection{Feynman rules}

Following standard procedure, we calculate the Feynman propagator for the scalar field with a minimum length.
Specifically, from the modified KG equation in Eq.\eqref{EoM} we have:
%
\begin{equation}\label{GreenFunc}
     \left[\partial_\mu\partial^\mu \left(1+ \gamma \partial_\nu \partial^\nu\right)^2 + (mc)^2 \right] G(x-x') = - i \delta(x-x')\,.
\end{equation}
Expressing the Green function $G(x - x')$ in terms of its Fourier transform
\begin{equation}\label{GreenFourier}
    G(x-x')=\int \frac{d^4p_0}{(2\pi)^4} \tilde G(p_0) e^{-i p_0\cdot(x-x')},
\end{equation}
and substituting it in Eq.\eqref{GreenFunc}, we get
\begin{equation}
\int \frac{d^4p_0}{(2\pi)^4} \tilde G(p_0) \left[-p_0^2 (1 - \gamma p_0^2)^2 + (mc)^2\right] e^{-i p_0\cdot(x-x')}=-i \int\frac{d^4p_0}{(2\pi)^4}e^{-i p_0\cdot(x-x')}\,.
\end{equation}
Therefore, the Fourier transform of the Feynman propagator has the form 
\begin{equation}
     \tilde G(p_0)  =\frac{-i}{-p_0^2 (1+ \gamma p_0^2)^2 + (mc)^2}\,,
\end{equation}
while the propagator itself is
\begin{equation}
     G(x-x')=\int \frac{d^4p_0}{(2\pi)^4} \frac{-i}{-p_0^2 ( 1 + \gamma p_0^2)^2 + (mc)^2}e^{-i p_0\cdot(x-x')}\,.
\end{equation}
The gauge field propagator can be treated in a similar manner.
In this case, the Feynman propagator has the following form
\begin{equation}
G(x-x')=\int \frac{d^4q_0}{(2\pi)^4} \frac{-i}{-q_0^2+2\gamma  q_0^4}e^{-i q_0\cdot(x-x')}\,,
\end{equation}
where $q_0$ is the auxiliary four-momentum of the gauge field.\\
To complete the Feynman rules for the system, we need to calculate the vertices for charged fields. Starting from the Lagrangian in Eq.\eqref{complexL}, we introduce 
the minimal coupling  
\begin{equation}
     \partial_\mu \rightarrow D_\mu=\partial_\mu -ie A_\mu\,,
\end{equation}
where $A_\mu$ is the gauge field.
Then, the full action of the minimally coupled 
complex scalar field and the gauge field reads
%
%
\begin{multline}
    \int \mathcal{L}\,d^4x = \int\left[\mathcal{L}_{A} + \mathcal{L}_{\phi,\mathbb{C}}\right]\,d^4x = \int\left\{\frac{1}{2} \left(D_{\mu} \phi\right)^\dagger D^{\mu} \phi - \frac{1}{2} m^2 \phi^\dagger \phi - \frac{1}{4}F^{\mu\nu}F_{\mu\nu}\right.\\
    + \gamma \left. \left[\left(D_\nu D^\nu D^\mu \phi\right)^\dagger D_\mu \phi + D_\nu D^\nu D^\mu \phi \left(D_\mu \phi\right)^\dagger \right] \right\}
%
\,d^4x\, .
\end{multline}
By expanding the covariant derivatives, we can write the Lagrangian as
\begin{multline}\label{covariantL}
    \mathcal{L} = \frac{1}{2} \left(\partial_{\mu} \phi\right)^\dagger \partial^{\mu} \phi - i e A^\mu \left[\phi^\dagger \partial_\mu \phi - \phi \left(\partial_\mu \phi\right)^\dagger \right] - \frac{1}{2} m^2 \phi^\dagger \phi + e^2 A_\mu A^\mu \phi^\dagger \phi - \frac{1}{4} F^{\mu\nu}F_{\mu\nu}\\
    \gamma \left\{\left(\partial_\nu \partial^\nu \partial^\mu \phi\right)^\dagger \partial_\mu \phi + \partial_\nu \partial^\nu \partial^\mu \phi \left(\partial_\mu \phi\right)^\dagger - \frac{1}{4} F^{\mu\nu} F_{\mu\nu} \partial_\mu \partial_\nu F^{\mu\nu} \right. \\
    - i e \left\{\left(\partial_\nu \partial^\nu A^\mu\right) \left[\phi^\dagger \partial_\mu \phi - \phi \left(\partial_\mu\phi\right)^\dagger \right] + 2 \partial_\nu A^\mu \left[\left(\partial^\nu \phi\right)^\dagger \partial_\mu \phi - \partial^\nu \phi \left(\partial_\mu \phi^\dagger\right)\right] \right.\\
    \left. + A^\mu \left[\left(\partial_\nu \partial^\nu \phi\right)^\dagger \partial_\mu \phi - \partial_\nu \partial^\nu \phi \left(\partial_\mu \phi^\dagger\right)\right]\right\}
   + e^2 \left\{ A^\nu(\partial_\nu A_\mu) \left[\left(\partial^\mu \phi\right)^\dagger \phi + (\partial^\mu \phi) \phi^\dagger\right]\right.\\ + 2A^\nu A_\mu \left[\left(\partial^\mu \phi\right)^\dagger \partial_\nu \phi + \left(\partial^\mu \phi\right) \left(\partial_\nu \phi\right)^\dagger\right] \\
    + A^\nu A_\nu \left[ \left(\partial^\mu \phi\right)^\dagger \partial_\mu \phi + \left(\partial^\mu \phi\right) \left(\partial_\mu \phi\right)^\dagger\right] - 2 A^\mu A^\nu \left[\phi^\dagger\partial_\nu\partial_\mu\phi + \phi \left(\partial_\nu\partial_\mu\phi\right)^\dagger\right]\\
    \left. + 2 A^\mu (\partial_\nu \partial^\nu A_\mu) \phi^\dagger \phi + A^\mu (\partial^\nu A_\mu) \left[\phi^\dagger \partial_\nu \phi  + \phi \left(\partial_\nu \phi\right)^\dagger\right] + A^\mu A_\mu \left[\phi^\dagger \partial_\nu \partial^\nu \phi + \phi \left(\partial_\nu \partial^\nu \phi\right)^\dagger\right]\right\}\\
    + i e^3 \left\{ A_\mu A_\nu A^\nu \left[\phi \left(\partial^\mu \phi\right)^\dagger - \phi^\dagger\partial^\mu\phi\right] + 2 A^\mu A_\mu A^\nu \left[\phi^\dagger \partial_\nu \phi - \phi \left(\partial_\nu \phi\right)^\dagger\right] \right\} \\
    \left. + 2 e^4 A^\mu A_\mu A^\nu A_\nu\phi^\dagger\phi\right\} 
    + \mathcal{O}(\gamma^2)\,.
\end{multline}
The above expression contains all terms corresponding to Feynman diagrams predicted by the usual scalar Quantum Electrodynamics (QED) Lagrangian with the addition of RGUP corrections. We see that up to  6-point vertices are allowed. This can be seen by examining  Eq.\eqref{covariantL} and observing that the maximum number of lines meeting at a vertex will have two scalars and four gauge bosons.
%
However, we will focus on the 3-point vertices, containing up to first order in the coupling constant $e$ and the RGUP coefficient $\gamma$.
We can use this approximation because the 3-point vertices will have the largest contribution to the scattering amplitudes.

\section{Corrections to the scalar QED scattering amplitudes}\label{sec:Crosssection}
The Lagrangian and the Feynman rules derived in the previous section deal with complex scalar fields, therefore they apply to charged systems that have no spin.
In this section, we will focus on calculating the corrections to the amplitudes of electromagnetic scattering of a scalar electron and a scalar muon, following \cite{Halzen:1984mc}.
The leading order of the scattering amplitude is provided by the three particle Feynman vertices derived from Eq.\eqref{covariantL}.
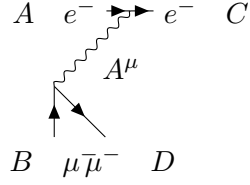
\begin{figure}
\feynmandiagram [small,baseline=(b.base),vertical=b to a] {
i1 [particle=\(A\quad e^{-}\)] -- [fermion] a -- [fermion] i2 [particle=\(e^{-}\quad C\)],
a -- [photon, edge label=\(A^\mu \)] b,
f1 [particle=\(B \quad\mu^{-} \)] -- [fermion] b -- [fermion] f2 [particle=\( \mu^{-}\quad D\)],
8};
\caption{The Feynman diagram of the scalar electron, scalar muon  scattering. }\label{muonscattering} 
\end{figure}
%
The transition amplitude for this case is given by integrating all three particle interaction terms 
the Eq.\eqref{covariantL}
\begin{equation}
    T_{fi}=-i\int  A^\mu j_\mu\,d^4x\,, \label{amplitude}
\end{equation}
where $j_{\mu}$ is the current corresponding to the electrons and muons.
In what follows, we will consider only terms up to first order in $\gamma$. The form of the transition amplitude shown in Eq.\eqref{amplitude} is the same for both vertices, where the potential is provided by the same gauge boson.
We then see that from Eq.\eqref{covariantL}, the terms containing one gauge field and two scalar fields will contribute to the following two terms in the transition amplitude
%
%
%
\begin{subequations} \label{amplitude_separated}
\begin{align}
 T_{fi}^{(1)}=&-i \int e A^\mu\left[\phi_f^\dagger \partial_\mu\phi_i-\phi_f\partial_\mu\phi_i^\dagger \right] d^4x\,,\\
 T_{fi}^{(2)}=&-i \int e\gamma  A^\mu\left[4\partial_\nu\partial^\nu\phi^\dagger \partial_\mu\phi+4\partial^\nu\phi^\dagger \partial_\nu\partial_\mu\phi+\phi^\dagger \partial_\nu\partial^\nu\partial_\mu\phi-4\partial_\nu\partial^\nu\phi\partial_\mu\phi^\dagger \right. \nonumber\\ 
 &\left.-4\partial^\nu\phi\partial_\nu\partial_\mu\phi^\dagger -\phi\partial_\nu\partial^\nu\partial_\mu\phi^\dagger  \right]d^4x\,.
\end{align}
\end{subequations}
The full transition amplitude for the three particle scattering is then given by the sum of these two terms
\begin{equation}
    T_{fi}= T_{fi}^{(1)}+ T_{fi}^{(2)}
    \,.
\end{equation}
The term $T_{fi}^{(1)}$, obtainable from the unmodified scalar QED Lagrangian, corresponds to the limit $\gamma \rightarrow 0$ of the expression above.
Comparing Eqs. \eqref{amplitude} and \eqref{amplitude_separated}, we can decompose the current $j_{\mu}$ into the unmodified  term and the RGUP correction
\begin{subequations} \label{currents}
\begin{align}\label{currents1}
 j_{fi\,\mu}^{(1)}=&-i e\left[\phi_f^\dagger \partial_\mu\phi_i-\phi_f\partial_\mu\phi_i^\dagger \right]\,,\\
\nonumber j_{fi\,\mu}^{(2)}=&-i e\gamma  \left[4\partial_\nu\partial^\nu\phi^\dagger \partial_\mu\phi+4\partial^\nu\phi^\dagger \partial_\nu\partial_\mu\phi+\phi^\dagger \partial_\nu\partial^\nu\partial_\mu\phi-4\partial_\nu\partial^\nu\phi\partial_\mu\phi^\dagger \right.\\ &\left.-4\partial^\nu\phi\partial_\nu\partial_\mu\phi^\dagger -\phi\partial_\nu\partial^\nu\partial_\mu\phi^\dagger\right]~.
\end{align}
\end{subequations}

Let us now assume the following standard form for the scalar field
\begin{equation}\label{Field}
    \phi(x_\mu)=N e^{-i p_\mu x^\mu}\,,
\end{equation}
where $N$ is the normalization constant and $p_\mu$ is the physical momentum of the field.
Using Eqs. \eqref{currents}, we obtain
\begin{subequations} \label{curr}
\begin{align}\label{curr1}
 j_{fi\,\mu}^{(1)}=&- e N_f N_i(p_f+p_i)_\mu e^{i(p_f-p_i)\cdot x}\,,\\
 j_{fi\,\mu}^{(2)}=&- e \gamma N_f N_i \left[-(p_f\cdot p_f)(4 p_i+-p_f)_\mu +4(p_f\cdot p_i)( p_i+p_f)_\mu\right.\nonumber\\
 &\left.-(p_i\cdot p_i)( p_i+4 p_f)_\mu\right]e^{i(p_f-p_i)\cdot x}
 \,.\label{curr4}
\end{align}
\end{subequations}

The Feynman diagram for the scattering is obtained by connecting two 3-point vertices through the gauge boson $A^\mu$, as shown in Fig. \ref{muonscattering}.
Thus we can express the scattering amplitude as 
\begin{equation}
  T_{ABCD}=-i \int  \left(j_{AC\,\mu}^{(1)}+ j_{AC\,\mu}^{(2)}
  \right)\frac{ig^{\mu\nu}}{-q+2\gamma q^4}
  \left(j_{BD\,\nu}^{(1)}+ j_{BD\,\nu}^{(2)}
  \right)\,.
\end{equation}
Using Eqs. \eqref{curr} and doing the integration we find an expression for the scattering amplitude
\begin{equation}\label{transition}
     T_{ABCD}=-N_A N_B N_C N_D (2\pi)^4 \delta^{(4)}(p_D+p_C-p_A-p_B)\mathfrak{M}\,,
\end{equation}
where $\delta^{(4)}(p_D+p_C-p_A-p_B)$ ensures conservation of momentum, $\mathfrak{M}$ is the invariant amplitude
\begin{multline}\label{invariantamplitude}
    -i \mathfrak{M}=-i \frac{e^2}{-q^2(1-2\gamma q^2)}\left\{(p_A+p_C)\cdot(p_B+p_D)\left[1-4\gamma \left(p_A\cdot p_C+p_B\cdot p_D\right)\right]\right.\\+\gamma (p_A+ p_C)\cdot\left(4p_B-p_D\right)m_{\mu^-}^2-\gamma\left(p_A+ p_C\right)\cdot\left(4p_D-p_B\right)m_{\mu^-}^2\\
    +\left.\gamma (p_B+ p_D)\cdot\left(4p_A-p_C\right)m_{e^-}^2-\gamma(p_B+ p_D)\cdot\left(4p_C-p_A\right)m_{e^-}^2\right\}\,,
\end{multline}
and the $\frac{-i}{-q^2+2\gamma q^4}$ is the propagator of the gauge field $A_\mu$. We omitted the terms containing higher than linear order in the RGUP coefficient $\gamma$.

Computing the scattering amplitude requires fixing the normalization for the free scalar field described by Eq.\eqref{covariantL}.
The temporal part of the current $j_\mu$ is the probability density while the spatial part the probability current density
\begin{equation}
    j^\mu=(\rho,\vec{\mathbf{j}})\,.
    \end{equation}
As per convention, we normalize the field such that the integral of the density over a fixed volume is equal to the sum of the energies of all the scalar particles in the system.
For the free field we thus have
\begin{align}\label{volume}
    \int_{V}\rho dV = & 2E\,, & \int_{V} \vec{j}\cdot d\vec{V} = & 2p\,.
\end{align}
From the modified KG equation we find the different components of the flux
\begin{align}\label{probdencitiy}
  \nonumber  \rho=&-i\left[\left(\phi^\dagger \frac{\partial\phi}{\partial t}-\phi \frac{\partial\phi^\dagger}{\partial t}\right)+2\gamma \frac{\partial}{\partial t}\left(\phi^\dagger \frac{\partial^2\phi}{\partial t^2}-\phi \frac{\partial^2\phi^\dagger}{\partial t^2}\right)\right.\\
    &\left.-4\gamma \left(\frac{\partial\phi^\dagger}{\partial t} \frac{\partial^2\phi}{\partial t^2}-\frac{\partial\phi}{\partial t} \frac{\partial^2\phi^\dagger}{\partial t^2}\right)+4\gamma \left(\phi^\dagger\nabla^2 \frac{\partial\phi}{\partial t}-\phi\nabla^2 \frac{\partial^2\phi^\dagger}{\partial t^2}\right)\right]\,,\\
    \nonumber  \vec{j}=-&i\left[\left(\phi^\dagger \nabla\phi-\phi\nabla\phi^\dagger\right)+4\gamma \left(\frac{\partial\phi^\dagger}{\partial t}\nabla \frac{\partial\phi}{\partial t}-\frac{\partial\phi}{\partial t}\nabla \frac{\partial^\dagger\phi}{\partial t}\right)\right.\\
    &\left.-2\gamma \nabla\left(\phi^\dagger\nabla^2\phi-\phi\nabla^2\phi^\dagger\right)+4\gamma \left(\nabla\phi^\dagger\nabla^2 \phi-\nabla\phi\nabla^2 \phi\right)\right]\label{momdensity}\,.
\end{align}
Using Eq.\eqref{Field}, Eq.\eqref{momdensity}, and  substituting in   Eq.\eqref{volume} we obtain the normalization constant
\begin{equation}\label{volumescale}
    N=\frac{1}{\sqrt{V\left(1-4\gamma (E^2+|\vec{p}|^2)\right)}}\,.
\end{equation}
Note that RGUP affects the normalization constant for the fields,  in addition to the corrections to the invariant amplitude $\mathfrak{M}$.

The scattering transition rate per unit volume is given by 
\begin{equation}
    W_{fi}=\frac{|T_{ABCD}|^2}{\tau V}\,,
\end{equation}
where $\tau$ is the time of interaction and $T_{ABCD}$ is the scattering transition amplitude.
Substituting Eq.\eqref{transition} with Eq.\eqref{volumescale} in the expression above, we have 
\begin{equation}
    W_{fi}=(2\pi)^4\frac{\delta^{(4)}(p_D+p_C-p_A-p_B)|\mathfrak{M}|^2}{\mu_A \mu_B \mu_C \mu_D \,V^4}\,,
\end{equation}
where
\begin{equation}
 \mu_A=1-4\gamma (E_A^2+|\vec{p}_A|^2)\,, \label{eqn:correction}
\end{equation}
and $\mu_B, \mu_C, \mu_D$ are similarly defined. 
Dividing the transition amplitude by the initial flux, and subsequently multiplying it by the number of final states in the volume, we get the cross section of the scattering process
\begin{equation}
    d\sigma=\frac{V^4}{|v_A|^2 2E_A 2E_B}\frac{\delta^{(4)}(p_D+p_C-p_A-p_B)|\mathfrak{M}|^2}{\mu_A \mu_B \mu_C \mu_D \,V^4}\frac{(2\pi)^4}{(2\pi)^6}\frac{d^3p_C}{2E_C}\frac{d^3p_D}{2E_D}\,,
\end{equation}
where $v_A=p_A/E_A$. We can rewrite the cross section  as 
\begin{equation}\label{eq.42}
    d\sigma=\frac{|\mathfrak{M}|^2}{F\,\mu_A \mu_B \mu_C \mu_D}dQ\,,
\end{equation}
where $F$ is the initial flux 
\begin{equation}\label{initialflux}
    F=|v_A|^2\, 2E_A \, 2E_B = 4\left(\left(p_A\cdot p_B\right)^2-m_A^2m_B^2\right)^{1/2}\,,
\end{equation}
and the Lorentz invariant phase space factor is 
\begin{equation}\label{LIPSF}
    dQ=\delta^{(4)}(p_D+p_C-p_A-p_B)\frac{(2\pi)^4}{(2\pi)^6}\frac{d^3p_C}{2E_C}\frac{d^3p_D}{2E_D}.
\end{equation}

\subsection{Application for high energy electron-muon scattering}

Let us consider a electron-muon scattering in the center of mass at very high energies.
In this case, the conservation of momentum reads
\begin{equation}\label{CM1}
    p_A+p_C=p_B+p_D=0\,.
\end{equation}
Notice that in this particular reference frame all the information is in the magnitudes of the initial and final momenta, $p_i$ and $p_f$, respectively, with
\begin{align}\label{CM2}
    |\vec{p}_A| = |\vec{p}_B| = & |\vec{p}_i|\,, & |\vec{p}_C| = |\vec{p}_D| = &|\vec{p}_f|\,.
\end{align}
In addition, for high energies we can consider the following approximation
\begin{equation}\label{ultrarelativistic}
    E^2\approx |\vec{p}|^2\,.
\end{equation}
Thus, we have for the correction terms $\mu$ in Eq.\eqref{eqn:correction}
\begin{align}
    \mu_A = \mu_B = & \mu_i, & \mu_C = \mu_D = & \mu_f\,.
\end{align}
Substituting them in Eq.\eqref{eq.42} gives for the differential cross section
\begin{equation}\label{CMscatteringamplitude}
d\sigma=\frac{|\mathfrak{M}|^2}{F\mu_i^2 \mu_f^2}dQ\,.
\end{equation}
Imposing conditions Eqs.(\ref{CM1},\ref{CM2},\ref{ultrarelativistic}) on Eq.\eqref{LIPSF}, the Lorentz invariant phase space factor $dQ$ can be expressed in terms of the solid angle $d\Omega$ as follows,
\begin{equation}\label{formfactor}
    dQ=\frac{1}{4\pi^2}\frac{|\vec{p}_f|}{4\sqrt{s}}d\Omega\,,
\end{equation}
where $s=(E_A+E_B)^2$ is the Mandelstam variables.
Substituting Eq.\eqref{formfactor} and Eq.\eqref{initialflux} in Eq.\eqref{CMscatteringamplitude} we get
\begin{equation}
    \left.\frac{d\sigma}{d\Omega}\right\vert_{CM}=\frac{1}{64\pi^2\,s\,\mu_i^2 \mu_f^2}\frac{|\vec{p}_f|}{|\vec{p}_i|}|\mathfrak{M}|^2\,.
\end{equation}
Imposing Eqs.(\ref{CM1},\ref{CM2},\ref{ultrarelativistic}) on Eq.\eqref{invariantamplitude}  and expanding the scalar product, we get the invariant amplitude for high energy collision in the center of mass frame $\mathfrak{M}$ up to first order in $\gamma$
\begin{equation}
    \mathfrak{M}=4\pi  \alpha \frac{3+\cos\theta}{1-\cos\theta}\left[1+8\gamma E^2(1-\cos\theta)\right]\,,
\end{equation}
where $\alpha$ is the fine structure constant.
%
%
\begin{figure}
 \centering
 \includegraphics[width=0.8\textwidth]{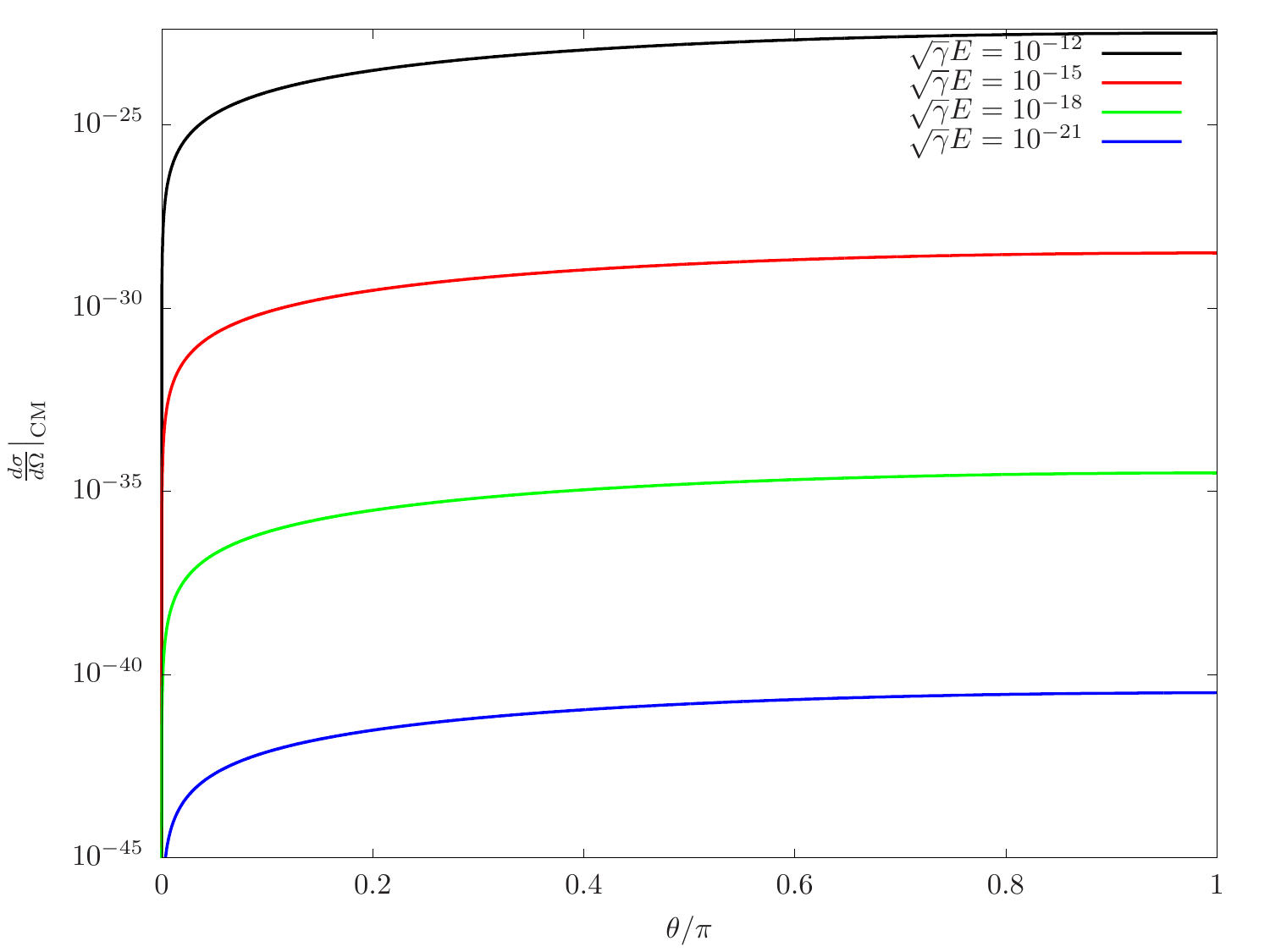}
\caption{Corrections to the differential cross section for different values of $\sqrt{\gamma} E$ with respect to the scattering angle $\theta$.
For the purposes of the plot, we consider the RGUP minimum length to be equal to the Planck length.}\label{fig2}
\end{figure}
Thus, the full expression for the differential cross section is 
\begin{equation}\label{final}
    \left.\frac{d\sigma}{d\Omega}\right\vert_{CM} = \frac{1}{4\,s}\alpha^2\left(\frac{3+\cos\theta}{1-\cos\theta}\right)^2\left[1+16\gamma E^2(1-\cos\theta)\right]\,.
\end{equation}
Notice that in the limit $\gamma \rightarrow 0$, we get the usual result, \emph{i.e.}
\begin{equation}
    \left.\frac{d\sigma}{d\Omega}\right\vert_{CM}^0 = \frac{1}{4\,s}\alpha^2\left(\frac{3+\cos\theta}{1-\cos\theta}\right)^2\,.
\end{equation}
Therefore, the magnitude of the correction is
\begin{equation}
    \frac{\left.d\sigma/d\Omega\right\vert_{CM} - \left.d\sigma/d\Omega\right\vert_{CM}^0}{\left.d\sigma/d\Omega\right\vert_{CM}^0} = 16\gamma E^2 (1 - \cos\theta).
\end{equation}
Finally, we notice that that correction is the largest for back scattering
($\theta=\pi$).
On Fig.\ref{fig2} one can see presented the RGUP correction term of Eq.\eqref{final} for several different energies. We assumed that the RGUP corrections will be relevant at Planck energies, {\it i.e.} the RGUP minimum measurable length is equal to the Planck length. The $\sqrt{\gamma}  E=10^{-16}$ curve represents the magnitude or the corrections for energies used in current particle physics experiments.  The curves $\sqrt{\gamma} E=10^{-14}$ and $\sqrt{\gamma}  E=10^{-12}$ correspond respectively to hundred and ten thousand times higher energies. We can easily see that even a simplified model such as this will give corrections to the cross section of the electromagnetic scattering. Moreover we can see that increasing the energy ten times will lead to a hundred times greater magnitude of the RGUP corrections.  Another feature worth mentioning is the fact that with the lowering of the energy the correction term is very quickly suppressed, and the modified theory recovers all previous results.

\section{Conclusion}\label{sec:Conclusion}
Summarizing, in a previous work \cite{Todorinov:2018arx}, we proved that the  Relativistic Generalized Uncertainty Principle and its corresponding Lorentz invariant minimum length leads to a modification of the dispersion relation and the Klein--Gordon equation. 
Using the Ostrogradsky method for higher derivative Lagrangians, we derived the Lagrangian for Scalar Quantum Electrodynamics with minimum length from the RGUP modified equation of motion.
We then derived Feynman propagators for the scalar and the gauge fields, in addition to Feynman interaction vertices.
We observed that the inclusion of Quantum Gravity corrections introduced three additional three-particle Feynman vertices and a host of others.
Using our results, we were able to derive the transition amplitudes for the scalar electron-muon scattering in Fig.\ref{muonscattering}.
We found that one of the consequences of minimum length is  corrections to the  leading order of the invariant amplitude. This can be seen from Eq.\eqref{invariantamplitude}.\\
Then we applied our results to a ultra-relativistic case, finding the differential cross-section Eq.\eqref{final} and its RGUP corrections.
These results will allow better understanding of the effects of Quantum Gravity.
Moreover, they pave the way towards calculating  Quantum Gravity  corrections to scattering processes involving  Quantum Electrodynamics. Additionally Eq.\eqref{RealL} is the RGUP modified Lagrangian for free real scalar field. One example of such a field is the Higgs field. Using this and the RGUP modified  gauge field langrangian  opens the way to doing Higgs mechanism in quantum field theories with minimum length. Furthermore one can calculate the corrections to the masses of the weak interaction bosons, or topological defects that might arise from the spontaneous symmetry breaking. \\
We also obtained a gauge invariant Quantum Gravity modified abelian gauge field Lagrangian, with small scale self-interactions between the bosons.\\
This will result in Quantum Gravity modified  Maxwell's equations, leading to interesting Quantum Gravity effects on processes like Faraday rotation.
Possibly, it can be measured by exploring the properties of Cosmic Background Radiation, using the long distance as an amplifier for the RGUP effects.

\noindent
{\bf Acknowledgement}

 This work was supported by the Natural Sciences and Engineering Research Council of Canada, and University of Lethbridge. We thank M. Fridman for discussions.


\appendix

\section{Equations of motion and the Lagrangian} \label{apx}

Following the Ostrogradsky method for higher derivative Lagrangians presented in \cite{deUrries:1998obu,Woodard:2015zca,Pons:1988tj}, Eq.\eqref{EoM} is obtained by applying the Euler-Lagrange equations to the Lagrangian.
In other words, the Lagrangian below 
\begin{align}\label{L}
   \nonumber \mathcal{L}=\frac{1}{2}\partial_{\mu}\phi\partial^{\mu}\phi&+\gamma \left( C_1\,\partial_\mu\partial^\mu\phi\,\partial_\nu\partial^\nu\phi+C_2\,\partial_\mu\phi\,\partial^\mu\partial_\nu\partial^\nu\phi+C_3\,\partial_\nu\partial^\nu\partial^\mu\phi\,\partial_\mu\phi\right)\\\nonumber&+\gamma^2\left(C_4\, \partial_\mu\partial^\mu\partial_\nu\phi\,\partial^\nu\partial_\rho\partial^\rho\phi+C_5\, \partial_\mu\partial^\mu\partial_\nu\partial^\nu\phi\,\partial_\rho\partial^\rho\phi+C_6\, \partial_\mu\partial^\mu\partial_\nu\partial^\nu\partial_\rho\phi\,\partial^\rho\phi\right.\\
   &\left.+C_7\, \partial_\mu\partial^\mu\phi\,\partial_\nu\partial^\nu\partial_\rho\partial^\rho\phi+C_8\, \partial_\mu\phi\,\partial^\mu\partial_\nu\partial^\nu\partial_\rho\partial^\rho\phi\right)+C_9m^2\phi^2\,,
\end{align}
which is the most general form of the Lagrangian, which has up to fourth order derivatives.
The application of the Ostrogradsky method to  Eq.\eqref{L}, should  give rise to the equations of motion in Eq.\eqref{EoM}.
According to the Ostrogradsky method, the Euler-Lagrange equations for theories with higher derivatives will have the form:
\begin{equation}
    \frac{dL}{dq} -\frac{d}{dt}\frac{dL}{d\dot{q}}+\frac{d^2}{dt^2}\frac{dL}{d\ddot{q}}+\ldots+(-1)^n\frac{d^n}{dt^n}\frac{dL}{d (d^nq/dt^n)}=0\,,
\end{equation}
which in the case of fields is
\begin{equation}
    \frac{\partial\mathcal{L}}{\partial\phi}-  \partial_\mu  \frac{\partial\mathcal{L}}{\partial(\partial_\mu\phi)}+   \partial_{\mu_1} \partial_{\mu_2}\frac{\partial\mathcal{L}}{\partial(\partial_{\mu_1} \partial_{\mu_2}\phi)}+\ldots
    +(-1)^m\partial_{\mu_1}\ldots\partial_{\mu_m}\frac{\partial\mathcal{L}}{\partial(\partial_{\mu_1} \ldots\partial_{\mu_m}\phi)}=0\,.
\end{equation}
We can now calculate the Euler-Lagrange equations for the Lagrangian Eq.\eqref{L}
\begin{align}
\nonumber& 2C_9 m^2c^2\phi-\partial_\mu\partial^\mu \phi +2\gamma C_2 \partial_\mu\partial^\mu\partial_\nu\partial^\nu\phi+2\gamma C_3 \partial_\mu\partial^\mu\partial_\nu\partial^\nu\phi-\gamma^2C_6\partial_\mu\partial^\mu\partial_\nu\partial^\nu\partial_\rho\partial^\rho\phi\\
\nonumber&-\gamma^2C_8\partial_\mu\partial^\mu\partial_\nu\partial^\nu\partial_\rho\partial^\rho\phi+4\gamma C_1\partial_\mu\partial^\mu\partial_\nu\partial^\nu\phi +\gamma^2C_5\partial_\mu\partial^\mu\partial_\nu\partial^\nu\partial_\rho\partial^\rho\phi+\gamma^2C_7\partial_\mu\partial^\mu\partial_\nu\partial^\nu\partial_\rho\partial^\rho\phi\\\nonumber&
+2\gamma C_2 \partial_\mu\partial^\mu\partial_\nu\partial^\nu\phi+2\gamma C_3\partial_\mu\partial^\mu\partial_\nu\partial^\nu-\gamma^2C_4\partial_\mu\partial^\mu\partial_\nu\partial^\nu\partial_\rho\partial^\rho\phi+\gamma^2C_5\partial_\mu\partial^\mu\partial_\nu\partial^\nu\partial_\rho\partial^\rho\phi\\ &+\gamma^2C_7\partial_\mu\partial^\mu\partial_\nu\partial^\nu\partial_\rho\partial^\rho\phi-\gamma^2C_7\partial_\mu\partial^\mu\partial_\nu\partial^\nu\partial_\rho\partial^\rho\phi-\gamma^2C_8\partial_\mu\partial^\mu\partial_\nu\partial^\nu\partial_\rho\partial^\rho\phi=0\,,
\end{align}
simplifying 
\begin{align}\nonumber
    2C_9m^2\phi&+\partial_\mu\partial^\mu\phi-4\gamma (C_2+C_3+C_1)\partial_\mu\partial^\mu\partial_\nu\partial^\nu\phi\\&+\gamma^2(C_6+2C_8-2C_5-C_7+C_4)\partial_\mu\partial^\mu\partial_\nu\partial^\nu\partial_\rho\partial^\rho=0\,,
\end{align}
which we compare to the Eq.\eqref{EoM}, and we get
\begin{align}
    &C_9=\frac{1}{2}\\
    &C_1+C_2+C_3=\frac{1}{2}\\
    &C_6+2C_4-C_7+2C_8-2C_5=1
    \end{align}

Simplifying  the resulting  Lagrangian term by term, to remove the surface terms, we get
\begin{align}
    C_1\,\partial_\mu\partial^\mu\phi\,\partial_\nu\partial^\nu\phi&=\underbrace{C_1 \partial_\mu\left(\partial^\mu\phi\,\partial_\nu\partial^\nu\phi\right)}_{surface}-C_1\partial^\mu\phi\,\partial_\mu\partial_\nu\partial^\nu\phi\\
  \nonumber  C_4\, \partial_\mu\partial^\mu\partial_\nu\phi\,\partial^\nu\partial_\rho\partial^\rho\phi&=C_4\partial_\nu\left(\partial_\mu\partial^\mu\phi\partial^\nu\partial_\rho\partial^\rho\phi\right)-C_4\partial_\mu\partial^\mu\phi\partial_\nu\partial^\nu\partial_\rho\partial^\rho\phi\\&=\underbrace{C_4\partial_\nu\left(\partial_\mu\partial^\mu\phi\partial^\nu\partial_\rho\partial^\rho\phi\right)}_{surface}-\underbrace{C_4\partial_\mu\left(\partial^\mu\phi\partial_\nu\partial^\nu\partial_\rho\partial^\rho\phi\right)}_{surface}+C_4\partial^\mu\phi\,\partial_\mu\partial_\nu\partial^\nu\partial_\rho\partial^\rho\phi\,
\end{align}
we repeat the process for all the terms.  We also use the fact that for scalar fields $[\phi,\phi]=0$, in other words the field is commutative. Therefore 
\begin{equation}
\partial^\mu\phi\,\partial_\mu\partial_\nu\partial^\nu\phi=\partial_\mu\partial_\nu\partial^\nu\phi\,\partial^\mu\phi\,.
\end{equation}
Applying this the Lagrangian allows us to combine the $C_1$, $C_2$, and $C_3$ terms into one $C$, in addition the $C_4\,,\ldots\,C_8$ can also be combined into one $D$.  The resulting Lagrangian is 
\begin{equation}
 \mathcal{L}=\frac{1}{2}\partial_{\mu}\phi\partial^{\mu}\phi+C_9m^2\phi^2+\gamma C\,\partial_\nu\partial^\nu\partial^\mu\phi\,\partial_\mu\phi+\gamma^2D\partial_\mu\phi\,\partial^\mu\partial_\nu\partial^\nu\partial_\rho\partial^\rho\phi\,.
\end{equation}
We again see that this is a higher derivative Lagrangian, therefore we need to use the Ostrogradsky method to calculate the equation of motion
\begin{equation}
     \frac{\partial\mathcal{L}}{\partial\phi}-  \partial_\mu  \frac{\partial\mathcal{L}}{\partial(\partial_\mu\phi)}-\partial_\mu \partial_\nu\partial^\nu \frac{\partial\mathcal{L}}{\partial(\partial_\mu\partial_\nu\partial^\nu\phi)}-\partial_\mu \partial_\nu\partial^\nu\partial_\rho\partial^\rho \frac{\partial\mathcal{L}}{\partial(\partial_\mu \partial_\nu\partial^\nu\partial_\rho\partial^\rho)}=0
\end{equation}
\begin{align}
  \nonumber  C_9m^2\phi-\partial_\mu\partial^\mu\phi-C\gamma \partial_\mu\partial^\mu\partial_\nu\partial^\nu\phi-C\gamma \partial_\mu\partial^\mu\partial_\nu\partial^\nu\phi\qquad\qquad&\\-D\gamma^2\gamma \partial_\mu\partial^\mu\partial_\nu\partial^\nu\partial_\rho\partial^\rho\phi-D\gamma^2\gamma \partial_\mu\partial^\mu\partial_\nu\partial^\nu\partial_\rho\partial^\rho\phi&=0\\
   C_9m^2\phi-\partial_\mu\partial^\mu\phi-2C\gamma \partial_\mu\partial^\mu\partial_\nu\partial^\nu\phi-2D\gamma^2\gamma \partial_\mu\partial^\mu\partial_\nu\partial^\nu\partial_\rho\partial^\rho\phi&=0\,.
\end{align}
Comparing the above to Eq.\eqref{EoM} we can fix the parameters 
\begin{align}
    C_9&=-\frac{1}{2}\\
    C&=-1\\
    D&=\frac{1}{2}\,.
\end{align}
We can easily see that the coefficients are unique.

\end{document}